\documentclass[fleqn,usenatbib]{mnras}

\usepackage{newtxtext,newtxmath}

\usepackage[T1]{fontenc}

\DeclareRobustCommand{\VAN}[3]{#2}
\let\VANthebibliography\thebibliography
\def\thebibliography{\DeclareRobustCommand{\VAN}[3]{##3}\VANthebibliography}

\usepackage{graphicx}	
\usepackage{amsmath}	
\usepackage{xcolor,ulem}

\newcommand{\Msun}{\ensuremath{\mathrm{M}_\odot}}

\title[Late-time radio observations of GRB\,200522A]{Late-time radio observations of the short GRB\,200522A: constraints on the magnetar model}

\author[G. Bruni et al.]{
G. Bruni,$^{1}$\thanks{E-mail: gabriele.bruni@inaf.it}
B. O'Connor,$^{2,3,4,5}$ 
T. Matsumoto,$^{6,7,8}$
E. Troja,$^{2,3}$
T. Piran,$^{6}$
L. Piro,$^{1}$
and
R. Ricci$^{9,10}$
\\
$^{1}$INAF -- Istituto di Astrofisica e Planetologia Spaziali, via Fosso del Cavaliere 100, I-00133 Roma, Italy\\
$^{2}$ Astrophysics Science 
Division, NASA Goddard Space Flight Center, 8800 Greenbelt Rd, Greenbelt, MD 20771, USA\\
$^{3}$ Department of Astronomy, University of Maryland, College Park, MD 20742-4111, USA\\
$^{4}$ Department of Physics, The George Washington University, 725 21st Street NW, Washington, DC 20052, USA\\
$^{5}$ Astronomy, Physics, and Statistics Institute of Sciences (APSIS), The George Washington University, Washington, DC 20052, USA\\
$^{6}$ Racah Institute of Physics, Edmund J. Safra Campus, Hebrew University of Jerusalem, Jerusalem 91904, Israel\\
$^{7}$ Research Center for the Early Universe, Graduate School of Science, University of Tokyo, Tokyo 113-0033, Japan\\
$^{8}$ Department of Physics, Graduate School of Science, University of Tokyo, Tokyo 113-0033, Japan\\
$^{9}$ Istituto Nazionale di Ricerche Metrologiche – Torino, Strada delle Cacce 91, I-10135 Torino, Italy\\
$^{10}$ INAF -- Istituto di Radioastronomia, via Gobetti 101, I-40129 Bologna, Italy\\
}

\date{Accepted XXX. Received YYY; in original form ZZZ}

\pubyear{2021}

\begin{document}
\label{firstpage}
\pagerange{\pageref{firstpage}--\pageref{lastpage}}
\maketitle

\begin{abstract}
GRB~200522A is a short duration gamma-ray burst (GRB) at redshift $z$=0.554 characterized by a bright infrared counterpart. A possible, although not unambiguous, interpretation of the observed emission is the onset of a luminous kilonova powered by a rapidly rotating and highly-magnetized  neutron star, known as magnetar. A bright radio flare, arising from the interaction of the kilonova ejecta with the surrounding medium, is a prediction of this model. Whereas the available dataset remains open to multiple interpretations (e.g. afterglow, r-process kilonova, magnetar-powered kilonova), long-term radio monitoring of this burst may be key to discriminate between models. We present our late-time upper limit on the radio emission of GRB~200522A, carried out with the Karl G. Jansky Very Large Array at 288 days after the burst. For kilonova ejecta with energy $E_{\rm ej}$\,$\approx$10$^{53}\,\rm erg$, as expected for a long-lived magnetar remnant, we can already rule out ejecta masses $M_{\rm ej}\lesssim0.03\,\Msun$ for the most likely range of circumburst densities $n\gtrsim 10^{-3}$ cm$^{-3}$. Observations on timescales of $\approx$\,3-10 yr after the merger will probe larger ejecta masses up to $M_{\rm ej}$\,$\sim$\,0.1\,$\Msun$, providing a robust test to the magnetar scenario.
\end{abstract}



\section{Introduction}

Short duration gamma-ray bursts (GRBs) are brief flashes of high-energy radiation produced, at least in some cases, by the coalescence of two neutron stars (NSs; \citealt{Eichler+1989}), as demonstrated by the association between the binary NS merger GW170817 and the short GRB 170817A \citep{Abbott17a, Abbott17b}. 
The fate of the merger remnant, either a massive NS or a black hole (BH), is still poorly constrained by either gravitational wave or electromagnetic observations
\citep{Pooley18, Ai18, Piro19,Abbott19, vanPutten19}, and remains a central question of the field with important implications for fundamental physics 
such as the nuclear equation of state (EoS).

After the merger, the lifetime of the surviving NS primarily depends on its mass and the unknown EoS of cold dense matter \citep{lp16}. Soft EoSs favor a prompt collapse to a BH, 
whereas stiffer EoSs allow for a dynamically stable NS \citep{piro17,GP13}. 
If the remnant NS is also highly magnetized, its large reservoir of rotational energy, up to 10$^{53}$ erg for a NS mass of 2.1 \Msun, could be transferred to the merger ejecta by the NS dipole spin-down radiation \citep{mb14}. 
These magnetar-driven winds accelerate the ejecta to transrelativistic speeds and substantially decrease its opacity by mixing \citep{Metzger+18}. Note also that any long-lived NS would decrease the opacity by neutrino irradiation \citep[e.g.][]{Metzger&Fernandez2014}.
When the ejecta undergo radioactive decays, the emitted signal, commonly referred to as either kilonova or macronova, is expected to peak at optical/UV energies within a few days after the merger, appearing bluer and brighter than a 
purely r-process powered kilonova \citep[e.g.][]{yu13,Wollager19}. 
On longer timescales, this fast ejecta interacts with the surrounding medium producing a luminous radio afterglow, visible  months to years after the merger \citep{NakarPiran11}. 
In some cases, radio emission from the nascent pulsar wind nebula may also become visible a few months after the merger \citep{PK13}.

Based on these models, events associated with bright and blue kilonovae are promising targets
to search for a late-time radio transient and test whether the luminosity of the optical and near-infrared (nIR) emission is the result of a long-lived magnetar. 
Past late-time radio surveys of short GRB locations all resulted in a non-detection,  thus disfavoring the presence of a stable magnetar in a large fraction of short GRBs \citep{Horesh16,Fong16,Klose19, Schroeder20,Ricci21,grandorf21}. The tightest upper limit for a short GRB with a kilonova candidate was derived for GRB~160821B \citep{Troja19,Lamb19}, 
and excluded ejecta masses $\lesssim$0.14\,M$_{\odot}$ for 
densities $n\gtrsim$10$^{-3}$\,cm$^{-3}$ and
an energy deposition of $E$\,$\approx$10$^{53}$\,erg 
\citep{Ricci21}. 
An upper limit $M_\textrm{ej}$\,$\lesssim$0.06\,M$_{\odot}$ was
also placed for GRB~150101B, a nearby short burst associated with a blue kilonova \citep{Troja18}. 
More recently, another event, GRB~200522A at redshift $z$=0.554, was associated with a luminous nIR counterpart
whose properties are open to multiple interpretations, including a bright kilonova emission. \citet{Fong21} suggested that the observed nIR signal could be the result of a magnetar-boosted kilonova, peaking at optical wavelengths. A bright late-time radio flare is therefore a natural expectation of this scenario. 

Additional observational evidence, presented by \citet{Oconnor21}, favored other interpretations. In particular, optical imaging simultaneous to the nIR observations does not detect a bright counterpart, whereas late-time X-ray observations find evidence of a long-lasting non-thermal emission. Overall, this evidence points to a substantial afterglow contribution up to late times. This may account for most, if not all, of the nIR signal if
a modest amount of dust is present along the line of sight. 
In any case, the observations of a bright, regular short GRB and afterglow modeling strongly suggest that GRB~200522A was viewed along its jet's axis. 
For this particular orientation, other factors may brighten the kilonova emission without invoking an additional energy source \citep{Heinzel21,Korobkin21}. 
A late-time radio flare may still be produced but it would peak at a later time and at a lower luminosity than in the magnetar scenario.

Recently we carried out radio observations of this event, obtaining a deep upper limit on radio emission almost a year after the event. 
Whereas the optical and nIR dataset is consistent with different interpretations,
we show here that late time radio observations are a suitable tool to determine whether a powerful magnetar has operated in this event and has boosted the nIR signal. We discuss the observations in \S \ref{sec:obs} and afterglow and kilonova constraints in \S \ref{sec:afterglow}. We describe  the model used to interpret the afterglow observations and the results  in \S \ref{sec:model}. We conclude with implications and the importance of further observations in \S \ref{sec:con}.


%

\section{Radio observations}
\label{sec:obs}
At early times, GRB~200522A was observed by \citet{Fong21} with the Karl G. Jansky Very Large Array (VLA) in C configuration. 
A weak radio counterpart was detected in C-band (6 GHz) a few hours after the burst and seen to fade below detection threshold within a few days. 
This radio signal may be produced either by a standard forward shock (FS) emission, a rapidly fading reverse shock (RS) component, or a combination of both. 
In any case, it was no longer detected at T$_0$+11~d with a 3$\sigma$ upper limit of 14 $\mu$Jy.

We re-observed the field with the VLA on February 19th, 2021 (T$_0$+288~d), with the array in A configuration (P.I. Troja). A single 2-hour run was carried out, applying a modified C-band (6 GHz) setup to avoid radio frequency interferences (RFI) from the Clarke belt satellites, and involving all the 28 antennas of the array. The net time on source was 01h22m. The source 3C\,147, observed at the beginning of the run, was used as both a bandpass and flux density scale calibrator,
while the phase calibrator paired to the target was J0022+0014. 
A phase-referencing cycle of 8 minutes, alternating between calibrator and target, was applied. 

Data were reduced with the the Common Astronomy Software Applications package ({\tt{CASA}}\footnote{\href{https://casa.nrao.edu/}{https://casa.nrao.edu/}}, \citealt{2007ASPC..376..127M}). We processed the raw data with VLA pipeline version 5.6.2-3. Imaging was performed with the {\tt{tclean}} task, applying natural weighting. This led to an angular resolution of 0.53$\times$0.36 arcsec. No emission was detected at the position of the radio afterglow detected by \cite{Fong21}. We measure an RMS noise of 2.6 $\mu$Jy/beam, resulting in a 3-$\sigma$ upper limit of 7.8 $\mu$Jy/beam. 
At a distance $z$\,$\sim$0.554 \citep{Oconnor21}, this corresponds to a luminosity
$L_{\nu}$\,$\lesssim$10$^{29}$\,erg\,s$^{-1}$\,Hz$^{-1}$ at 185~d (rest-frame) since the merger.


\begin{figure}
\begin{center}
\includegraphics[width=85mm, angle=0]{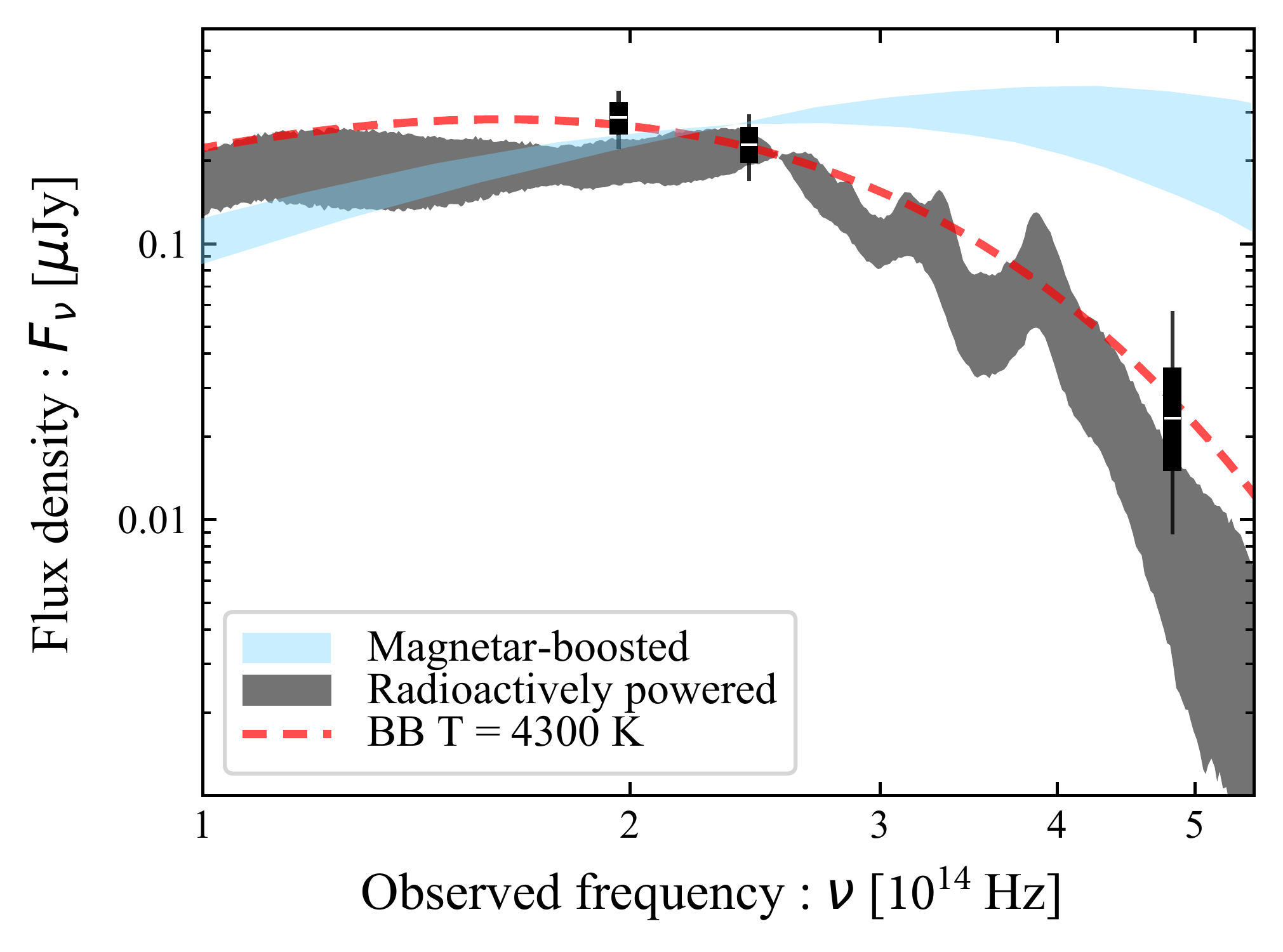}
\caption{
Flux density in the optical/nIR bands (after subtraction of the afterglow component) for GRB 200522A \citep{Oconnor21}.
Data are compared to the predicted emission from a magnetar-boosted kilonova (blue shaded region; \citealt{Fong21}), and a radioactively powered kilonova
from a lanthanide-poor ($Y_e$=0.37) wind ejected along the polar axis
(gray shaded region; \citealt{Oconnor21}). 
The red dashed line shows a simple blackbody (BB) spectrum with $T$\,$\sim$\,$4300$ K (rest frame). 
}
\label{BB SED}
\end{center}
\end{figure}


\section{Afterglow and kilonova constraints}
\label{sec:afterglow}
We use the afterglow models of \citet{Oconnor21} to bracket the range of possible circumburst densities. These models provide a good description of the broadband dataset, which includes 
X-ray data from \textit{Swift} and \textit{Chandra},  radio data from the VLA, optical photometry from Gemini, and nIR photometry  from the \textit{Hubble Space Telescope}. Since we wish to test the hypothesis of a luminous kilonova emission, we only consider models of a forward-shock (FS) plus black-body (BB) emission.

Multiple solutions can fit the available data equally well: $a$) a bright early-peaking FS expanding into a homogeneous medium with density $10^{-4}$\,cm$^{-3} \lesssim n \lesssim 10^{-1}$ cm$^{-3}$ (1\,$\sigma$) , and $b$) a bright reverse shock (RS) plus a FS  expanding in a medium with a slightly lower density $2\times10^{-5}$\,cm$^{-3} \lesssim n \lesssim 10^{-2}$ cm$^{-3}$.
In both cases, the total (beaming corrected) jet energy is $E_{\rm j}\approx4\times10^{49}\,\rm erg$. This set of solutions also includes the models presented in \citet{Fong21}, who fix the fraction of the electron energy to the total kinetic energy $\varepsilon_{\rm e}$=0.1 and the fraction of magnetic field energy $\varepsilon_{\rm B}$=0.01. 
Extremely low-density solutions with  $n\approx2\times10^{-5}\,\rm cm^{-3}$ appear unlikely given the location of the GRB within its host ($\sim1.1$ kpc from the host galaxy's nucleus; \citealt{Oconnor21,Fong21}), 
and we therefore focus our work on the range 
$10^{-4}$\,cm$^{-3} \lesssim n \lesssim 10^{-1}$ cm$^{-3}$ \citep{Oconnor21}.

A kilonova component, although not required, can be constrained by the data as shown in Fig. \ref{BB SED}. 
After subtracting the afterglow contribution, 
the residual optical and nIR emission can be modelled with a simple blackbody
with rest-frame temperature $T$\,$\sim$\,$4300$~K \citep{Oconnor21}. 
This is consistent with a model of a radioactively powered kilonova with large
ejecta mass $0.03\,M_{\odot} \lesssim M_\textrm{ej}\lesssim$0.1\,$M_{\odot}$ \citep{2018MNRAS.478.3298W,Korobkin21}.
In the magnetar-boosted scenario, 
comparably massive ejecta, $0.01\,M_{\odot} \lesssim M_\textrm{ej}\lesssim$0.1\,$M_{\odot}$, can reproduce the observed
nIR luminosity. However, 
the peak of the kilonova emission lies in the optical range ($T\sim6400-7000\,\rm K$, \citealt{Fong21}) and 
overpredicts the optical flux by a factor of $\gtrsim$\,$8$. 
 In order for the optical emission to be consistent with this model, an intrinsic extinction of $A_V$\,$\gtrsim$\,$1$ mag should be considered.

Due to the limited dataset, neither scenario can be firmly ruled out on the basis of the optical/nIR data alone. However, the predictions for a late-time radio flare differ greatly. In the magnetar model, a 
remnant NS with $M$\,$\sim$2.1\,\Msun\ and initial spin period
$P_0$\,$\sim$0.7~ms  will impart to the ejecta a large kinetic energy, 
$E_\textrm{ej}$\,$\approx$\,$10^{53}$ erg.  Whereas for a radioactively powered kilonova with ejecta mass $M_{\textrm{ej}}$\,$\sim$\,$0.03-0.1 M_\odot$ and velocity $v_{\textrm{ej}}$\,$\sim$\,$0.15c$, as presented by \citet{Oconnor21}, the 
energy is significantly smaller ($\approx$\,$10^{51}$~erg). Late-time radio observations can discern between these two scenarios as we discuss below.


\section{Implication of the late radio observations}
\label{sec:model} 
To test the magnetar-boosted model, we calculate the radio flare arising from the energized kilonova ejecta. The kilonova ejecta with a kinetic energy $E_{\rm ej}$ and mass $M_{\rm ej}$ propagates into an ISM with a density $n$. At the shock front, the magnetic field is amplified and electrons are accelerated to a power-law distribution, radiating synchrotron emission \citep{NakarPiran11}. Our calculation method follows \cite{Ricci21}. In this particular case, we do not consider the afterglow suppression by the relativistic jet \citep{Margalit&Piran2020}, whose timescale is much shorter ($\lesssim0.2\,\rm yr$) than the time of our observations.
Following the magnetar-boosted model presented in \citet{Fong21}, we focus mainly on a kinetic energy of $E_{\rm ej}=10^{53}\,\rm erg$.

\begin{figure}
\begin{center}
\includegraphics[width=85mm, angle=0]{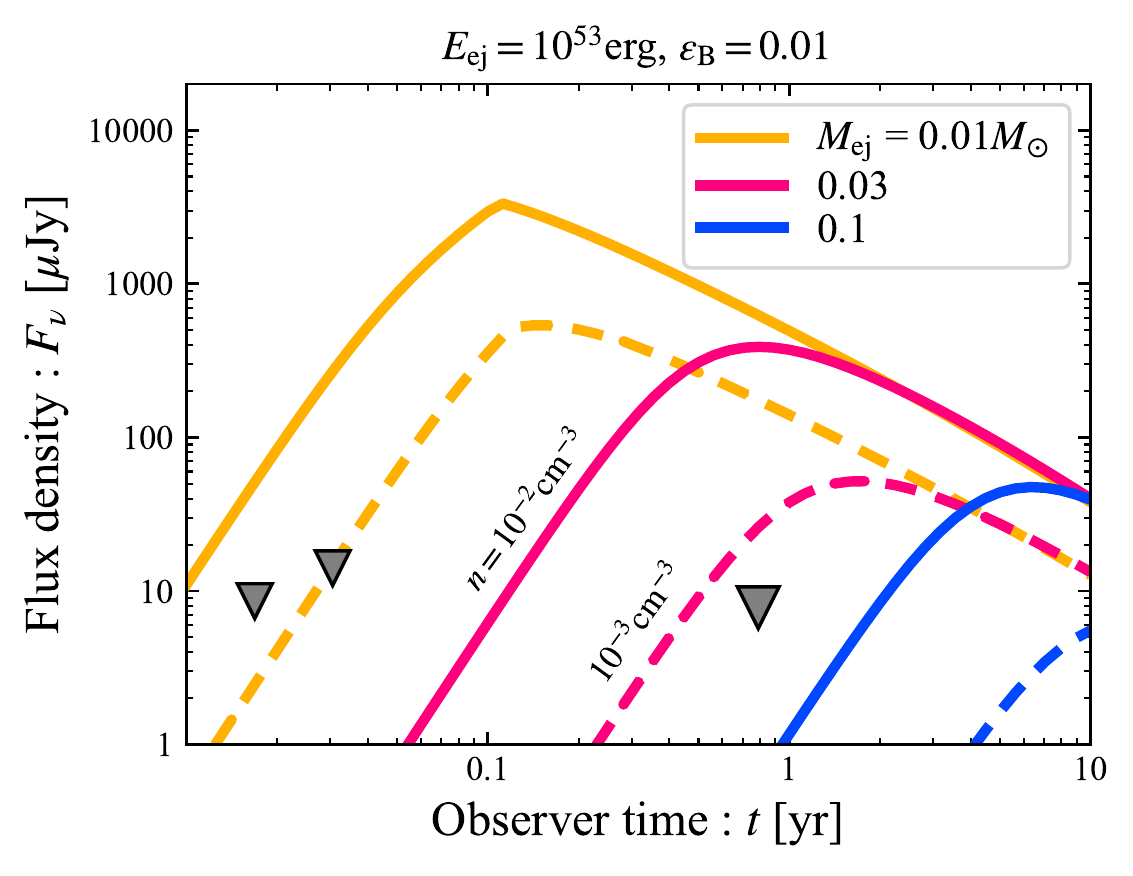}
\caption{Light curves of GRB 200522A for various ejecta masses of $M_{\rm ej}=0.01$, $0.03$, and $0.1\,\Msun$ (the corresponding initial Lorentz factors are $\Gamma_{\rm in}=6.6$, $2.9$, and $1.6$ or $\beta_{\rm in}=0.77$ for the last case). Solid and dashed lines correspond to densities $10^{-2}\,\rm cm^{-3}$ and $10^{-3}\,\rm cm^{-3}$, respectively.
The other parameters are $\varepsilon_{\rm e}=0.1$, $\varepsilon_{\rm B}=0.01$, and $p=2.5$. 
Gray triangles show 3 $\sigma$ flux density upper-limits 
from earlier observations by \citet{Fong21}, and our late-time monitoring}. 

\label{fig lc}
\end{center}
\end{figure}

Fig. \ref{fig lc} depicts the radio light curves for different ejecta masses of $M_{\rm ej}=0.01-0.1\,\Msun$, consistent with the kilonova modeling, and densities of $n=10^{-3}$ and $10^{-2}\,\rm cm^{-3}$, representative of the median values found through broadband afterglow fitting. We set the power-law index of the electron distribution $p=2.5$, and microphysical parameters $\varepsilon_{\rm e}=0.1$ and $\varepsilon_{\rm B}=0.01$ identical to \cite{Fong21}. 
We adopt these parameters throughout unless otherwise specified. The observed frequency, $6$ GHz, is typically larger than both the  synchrotron frequency $\nu_{\rm m}$ and the self-absorption frequency $\nu_{\rm a}$ resulting from the adopted parameters \citep[see][]{Ricci21}. The light curve peaks at the deceleration time, with more massive ejecta leading to a later peak and a lower peak flux. 


\begin{figure}
\begin{center}
\includegraphics[width=85mm, angle=0]{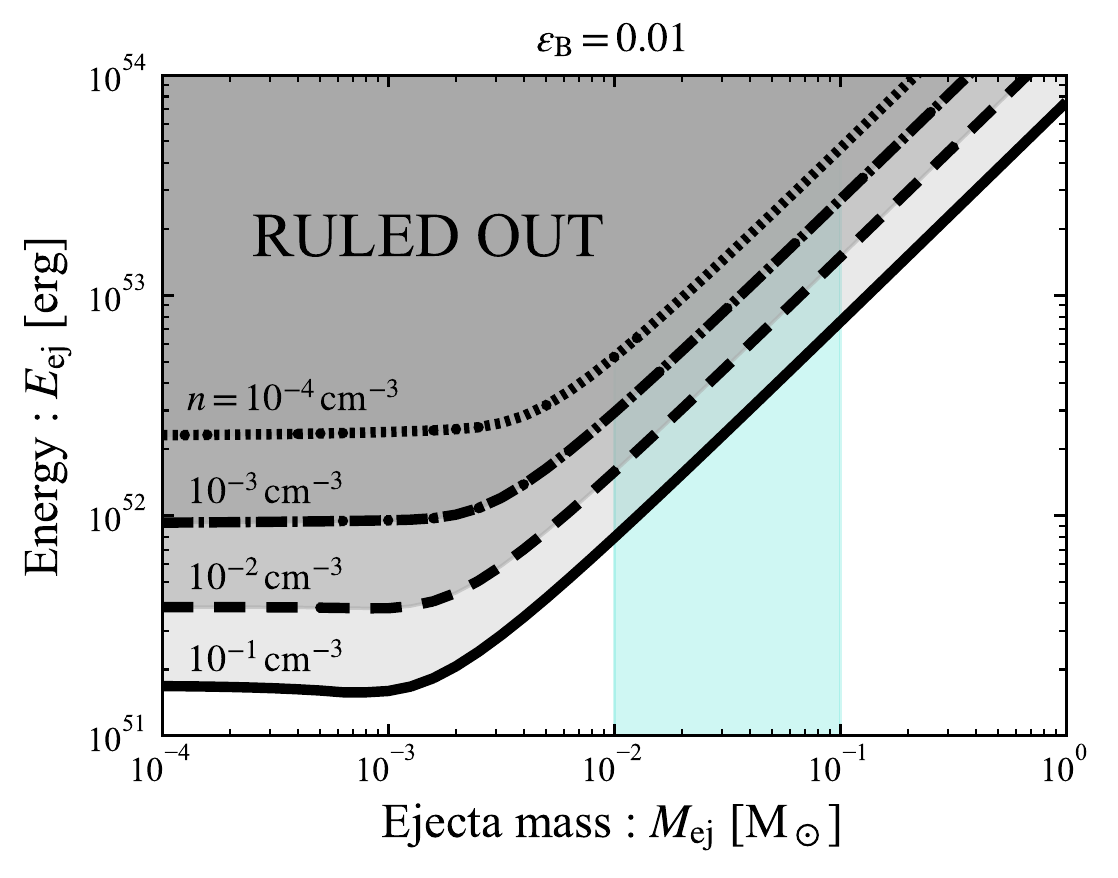}
\caption{Constraints on the kinetic energy for the allowed range of densities ($10^{-4}$ to $10^{-1}\,\rm cm^{-3}$) for GRB 200522A. For each density value, 
the range of energies lying above the curve is ruled out by the radio data.  The other parameters are fixed to $\varepsilon_{\rm e}=0.1$, $\varepsilon_{\rm B}=0.01$, and $p=2.5$. The blue shaded region represents the relevant range of ejecta mass.}
\label{fig const}
\end{center}
\end{figure}


By imposing that the flux $F_{\nu}$ is smaller than the observed upper limit, we derive a lower limit on the ejecta mass. 
With $E_{\rm ej}=10^{53}\,\rm erg$ the outflow is in the relativistic (or at least mildly relativistic) regime in which the flux depends most sensitively on the Lorentz factor $\Gamma$
and hence on the mass:
\begin{align}
F_{\nu}\simeq40{\,\rm \mu Jy\,}~\varepsilon_{\rm e,-1}^{p-1}~\varepsilon_{\rm B,-2}^{{(p+1)}/{4}}~n_{-3}^{({p+5})/{4}}~d_{28}^{-2}~\nu_{\rm GHz}^{({1-p})/{2}}~t_{\rm yr}^{3}~\bigg(\frac{\Gamma}{2}\bigg)^{2(p+3)},
    \label{eq flux}
\end{align}
where $d$, $\nu$, and $t$ are the distance (cm), observing frequency (GHz), and time since the merger (years), respectively. 
We use the notation $Q_x=Q/10^x$ in cgs units for the other quantities.
The upper limit on the observed flux gives an upper limit on the Lorentz factor which corresponds to a lower limit on the ejecta mass by $E_{\rm ej}\simeq\Gamma M_{\rm ej}c^2$:
\begin{align}
M_{\rm ej}\gtrsim0.03\,\Msun\,~E_{\rm ej,53}~\bigg[&\varepsilon_{\rm e,-1}^{\frac{p-1}{2}}~\varepsilon_{\rm B,-2}^{\frac{p+1}{8}}~n_{-3}^{\frac{p+5}{8}}\nonumber\\
&d_{28}^{-1}~t_{\rm yr}^{\frac{3}{2}}~\nu_{\rm GHz}^{\frac{1-p}{4}}~ F_{10\mu\rm Jy}^{-\frac{1}{2}}\bigg]^{\frac{1}{p+3}},
    \label{eq mass}
\end{align}
where $F_{10\mu \rm Jy}=F_{\nu}/10\,\mu\rm Jy$\,$\sim$0.78
is our limit.
Note that this mass limit depends linearly on the energy but it is almost independent from the uncertain parameters ($n$, $\varepsilon_{\rm e}$, and $\varepsilon_{\rm B}$) .

Fig. \ref{fig const} shows the parameter space of ejecta mass and energy ruled out by our upper limit of $7.8\,\mu\rm Jy$.
For an ejecta energy of $E_{\rm ej}$\,=\,$10^{53}$\,erg, the current limit reasonably rejects ejecta mass lower than $M_{\rm ej}\lesssim0.03\,\Msun$ for the most likely range of densities $n\gtrsim$10$^{-3}$\,cm$^{-3}$. Note that Eq. \eqref{eq mass} underestimates the mass limit by a factor of a few for $n=0.1\,\rm cm^{-3}$. This is because ejecta with $E_{\rm ej}\lesssim M_{\rm ej}c^2$ are in the Newtonian regime, so Eqs. \eqref{eq flux} and \eqref{eq mass} should be replaced with those for the Newtonian limit:
\begin{align}
F_{\nu}&\simeq150\,{\rm \mu Jy}~\varepsilon_{\rm e,-1}^{p-1}~\varepsilon_{\rm B,-2}^{{(p+1)}/{4}}~n_{-1}^{({p+5})/{4}}~d_{28}^{-2}~\nu_{\rm GHz}^{({1-p})/{2}}~t_{\rm yr}^{3}~\bigg(\frac{\beta}{0.8}\bigg)^{\frac{5p+3}{2}},\\
M_{\rm ej}&\gtrsim0.3\,\Msun\,~E_{\rm ej,53}~\bigg[\varepsilon_{\rm e,-1}^{\frac{p-1}{2}}~\varepsilon_{\rm B,-2}^{\frac{p+1}{8}}~n_{-3}^{\frac{p+5}{8}}d_{28}^{-1}~t_{\rm yr}^{\frac{3}{2}}~\nu_{\rm GHz}^{\frac{1-p}{4}}~ F_{10\mu\rm Jy}^{-\frac{1}{2}}\bigg]^{\frac{8}{5p+3}},
\end{align}
where $\beta$ is the ejecta velocity normalized by the speed of light. See also Eq. (26) in \citealt{Ricci21} for the deep-Newtonian case, where $\beta\lesssim0.2$.

\begin{figure}
\begin{center}
\includegraphics[width=85mm, angle=0]{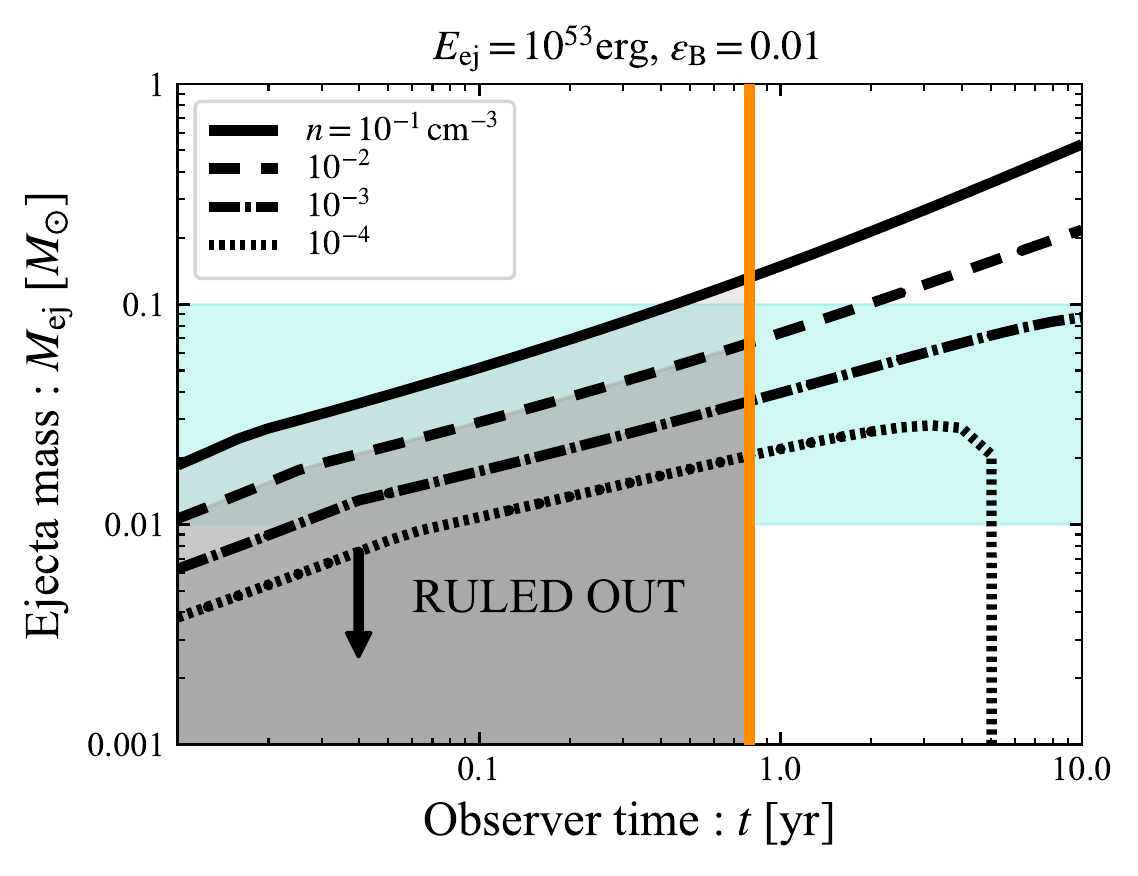}
\caption{Ejecta mass ruled out by null detection $F_{\nu}<7.8\,\mu \rm Jy$ as a function of time. For a fixed density, only values above each curve are allowed by the data.
The vertical orange line denotes the observation time in this work (288 days). The shaded region represents the relevant range of ejecta mass.}
\label{fig time}
\end{center}
\end{figure}

Later observations are needed to exclude more massive ejecta.  
Fig. \ref{fig time} depicts the ejecta mass ruled out by a similar null-detection as a function of time. Our limit can rule out $M_{\rm ej}\lesssim0.14\,\Msun$
for an environment denser than $n\sim$0.1\,cm$^{-3}$, which 
is consistent with the small projected offset from the galaxy's center
and the afterglow observations, although with low posterior probability ($\lesssim$15\%).
For lower densities of $n=10^{-2}$ and $n=10^{-3}\,\rm cm^{-3}$, 
favored by the afterglow data, 
observations in a few years will test ejecta of $0.03-0.1\,\Msun$.
An observation at 3 years after the merger would be effective
in excluding masses $M_{\rm ej}\lesssim0.1\,\Msun$ and 
$M_{\rm ej}\lesssim0.05\,\Msun$ for $n$\,=\,$10^{-2}$ and $n$\,=\,$10^{-3}\,\rm cm^{-3}$, respectively. 
For even lower densities $n<10^{-4}\,\rm cm^{-3}$ 
the peak flux falls below the expected upper limit
and late observations will not result in a significantly better constraint.

\begin{figure}
\begin{center}
\includegraphics[width=85mm, angle=0]{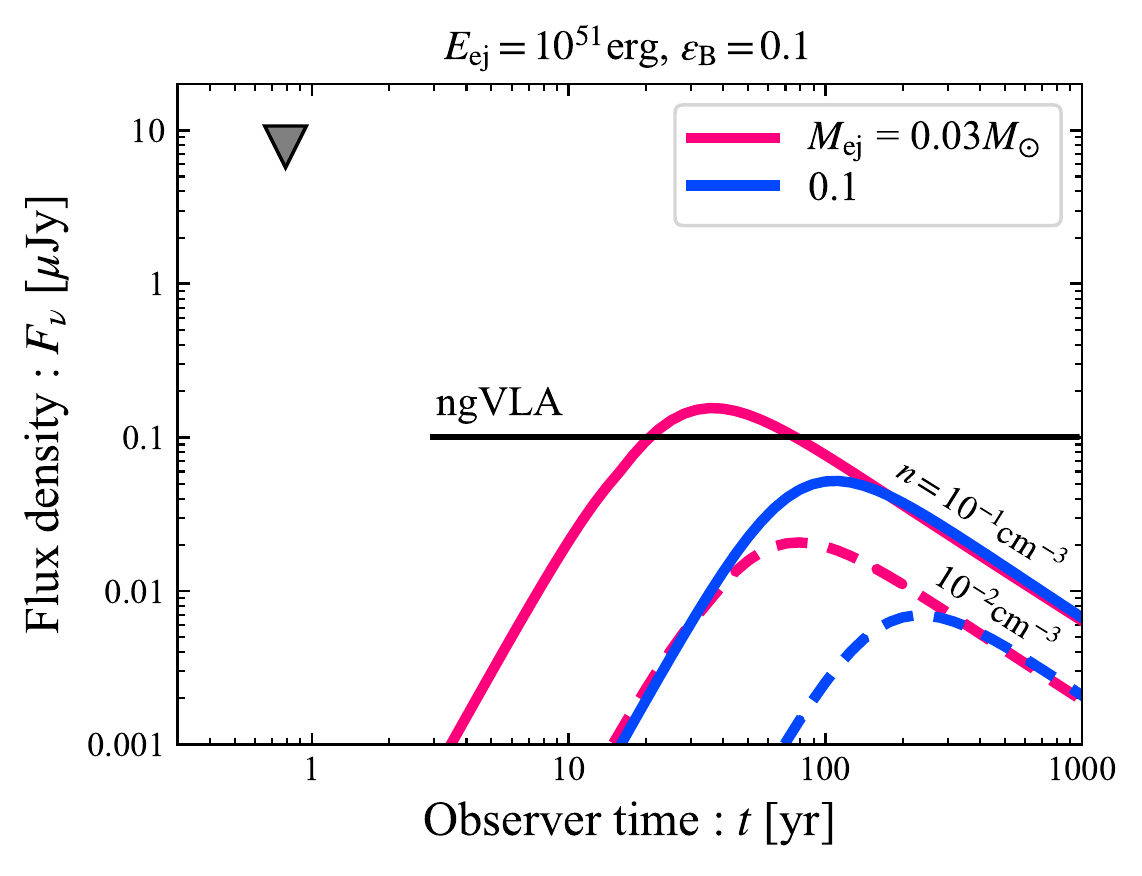}
\caption{The same as Fig. \ref{fig lc} but for the standard kilonova scenario ($E_{\rm ej}=10^{51}\,\rm erg$ and $M_{\rm ej}=0.03$ and $0.1\,\Msun$ corresponding to $\beta_{\rm in}=0.19$ and $0.11$, respectively). Black solid line represents the expected 3-$\sigma$ detection threshold of ngVLA in a $\sim$30~hr exposure.}
\label{fig lc2}
\end{center}
\end{figure}

Finally, we note that merger ejecta with kinetic energy of $E_{\rm ej}$\,$\sim$\,$10^{51}\rm \,erg$, as suggested by \cite{Oconnor21}, is allowed by the observations.
This is more explicitly demonstrated in Fig. \ref{fig lc2}.
Within this model, the peak timescale would be $\gtrsim30\,\rm yrs$ after the merger,
and the peak flux is $\lesssim0.2\,\mu \rm Jy$. Even with an optimistic choice of $\varepsilon_{\rm B}=0.1$ and the increased sensitivity of future observatories, the detection of a radio flare would be challenging for this burst. For example, considering the nominal RMS of 0.23 $\mu \rm Jy/beam$ in a 1-hour observation with the next generation VLA (ngVLA, \citealt{2018SPIE10700E..1OS}), 
it would take over 30 hrs to reach a 3-$\sigma$ detection (see also \citealt{2019BAAS...51c.209C} for an exhaustive discussion of radio counterparts studies with the ngVLA). This is expected in view of the large distance of this event as compared to those discussed in \cite{NakarPiran11}.


\section{Conclusions}
\label{sec:con}

We present VLA observations of the short duration GRB 200522A at 288 days post-merger. We do not detect a radio source at the GRB position, placing an upper limit of  $\nu L_{\nu}$\,$\lesssim$9$\times$10$^{38}$\,erg\,s$^{-1}$ at 9 GHz (rest-frame). This is comparable to the limits obtained for the larger sample of cosmological short GRBs at $z\lesssim$0.5 \citep{Ricci21,Schroeder20}.
In order to break the degeneracy in the interpretation of the GRB optical/nIR counterpart, we compare our radio limit to the theoretical predictions for a late-time radio flare produced by the merger ejecta. 

We first consider the scenario of a long-lived magnetar re-energizing 
the merger ejecta to $E_{\rm ej}=10^{53}\,\rm erg$.   
This model was invoked by \citet{Fong21} to interpret the 
nIR counterpart as a magnetar-boosted kilonova. 
Based on our VLA observation, we can exclude ejecta masses $M_{\rm ej}\lesssim0.03\,\Msun$ for a wide range of densities $n\gtrsim 10^{-3}$ cm$^{-3}$, also preferred by the GRB afterglow modeling \citep{Oconnor21}.

Additional late-time observations will provide even stronger constraints on the magnetar model.  
Assuming a similar sensitivity of $\approx$8 $\mu$Jy at 6~GHz, we find that 
future observations at $\sim$\,$3$ yr  and $\sim$\,$10$ yr post-merger 
would rule out $M_{\rm ej}\lesssim 0.1\,\Msun$ for $n$\,$\gtrsim$\,$10^{-2}\,\rm cm^{-3}$
and $n$\,$\gtrsim$\,$10^{-3}\,\rm cm^{-3}$, respectively.

Finally, we discuss the standard scenario of a radioactively powered kilonova,
also consistent with the optical/nIR dataset \citep{Oconnor21}. 
This model implies a significantly lower ejecta energy $E_{\rm ej}$\,$\approx$\,$10^{51}\,\rm erg$ as well as large masses,
0.03$\lesssim$\,$M_\textrm{ej}$\,$\lesssim$0.1 \Msun, 
and its associated radio transient would be too faint for detection by both current and future observatories. This is not unexpected in view of the distance of this object and the prediction for the late radio flares \citep{NakarPiran11}.
Events at much closer distance, such as the sample of short bursts
identified by \citet{dichiara20} and obviously GW170817 \citep{troja20,MAK20,Hajela21,troja21}, 
will be pivotal in constraining the late-time radio emission from
the merger ejecta.


\section*{Acknowledgements}
The authors thank the referee for useful comments that improved the manuscript.
The authors also thank Chris L. Fryer for providing the kilonova spectra used to model GRB 200522A.
E.T. and B.O. were supported in part by the National Aeronautics and Space Administration through grant GO21065A. T.P. was supported by an advanced ERC grant TReX. T.M. was supported by JSPS Postdoctral Fellowship, Kakenhi No. 19J00214 . LP acknowledges support from the European Union’s Horizon 2020 Programme under the AHEAD2020 project (grant agreement n. 871158) and from MIUR, PRIN 2017 (grant 20179ZF5KS).

\section*{Data Availability}
The data underlying this article will be shared on reasonable request to the corresponding author.



\bibliographystyle{mnras}







\bsp	
\label{lastpage}
\end{document}